\documentclass[
    reprint,
    superscriptaddress,
    amsmath,
    amssymb,
    aps,
    prb
]{revtex4-2}

\usepackage{array}
\usepackage{multirow}
\usepackage{makecell}
\usepackage{booktabs}
\usepackage{diagbox}
\usepackage{amsmath}
\usepackage{xcolor}
\usepackage{graphicx}
\usepackage{dcolumn}
\usepackage{bm}
\usepackage{gensymb}
\usepackage[normalem]{ulem}
\usepackage{orcidlink}

\definecolor{darkgreen}{RGB}{0,100,0}
\definecolor{darkpurple}{RGB}{100,40,140}

\begin{document}

%\preprint{APS/123-QED}

%\title{Phase Identification of CuInP$_2$S$_6$ through Raman and Infrared Vibrational Fingerprints}% Force line breaks with \\

%\title{Vibrational Fingerprints of Competing Structural and Polar Phases in CuInP$_2$S$_6$: A First-Principles Study}

\title{Vibrational Fingerprints of Competing Structural and Polar Phases in CuInP$_2$S$_6$}

\author{Pegah Mohammadi}
\affiliation{Department of Mechanical Engineering, University of Rochester, Rochester, New York 14627, USA}

\author{Sobhit Singh}
 \email{s.singh@rochester.edu}
\affiliation{Department of Mechanical Engineering, University of Rochester, Rochester, New York 14627, USA}
\affiliation{Material Science Program, University of Rochester, Rochester, New York 14627, USA} 

\date{\today}% It is always \today, today,

\begin{abstract}
%Layered two-dimensional van der Waals CuInP$_2$S$_6$ (CIPS) crystals have attracted significant attention due to their rich phase transitions under both temperature and pressure, which lead to diverse structural phases with distinctive physical properties. These phase transitions have been extensively investigated both experimentally and theoretically to identify the underlying phases and their characteristic behaviors. In this work, we employ first-principles density functional theory (DFT) calculations to study the vibrational properties of the reference paraelectric $C2/c$ and $P\overline{3}1c$ phases and their corresponding dynamically stable ferroelectric $Cc$ and $P31c$ phases, as well as the antiferroelectric $P\overline{1}$ and $P\overline{3}$ phases. By systematically analyzing the phonon spectra, we determine the Raman- and infrared-active modes for all candidate structures and identify distinct signature modes that enable reliable phase determination. Furthermore, using Porto's notation, we examine the polarization- and sample-orientation-dependent Raman selection rules under different scattering geometries and highlight the corresponding vibrational signatures.

Layered van der Waals CuInP$_2$S$_6$ crystals exhibit a rich sequence of temperature- and pressure-driven phase transitions, giving rise to multiple structural phases with distinct physical properties. 
Despite extensive experimental and theoretical efforts, unambiguous identification of these phases remains challenging because of their structural similarities and complex vibrational responses. 
In this work, we employ first-principles density functional theory calculations to systematically investigate the vibrational properties of the reference 
paraelectric $C2/c$ and $P\overline{3}1c$ phases and their corresponding dynamically stable ferroelectric $Cc$ and $P31c$ phases, as well as the antiferroelectric $P\overline{1}$ and $P\overline{3}$ phases.
From calculated phonon spectra and symmetry analysis, we determine the Raman- and infrared-active modes of each candidate structure and identify characteristic vibrational fingerprints that enable their phase determination through vibrational spectroscopy. 
Furthermore, using Porto notation, we establish polarization- and crystal-orientation-dependent Raman selection rules for different backscattering geometries and identify the modes that are expected to emerge or vanish under specific experimental configurations. 
Thus, our results provide a comprehensive theoretical framework for interpreting Raman and infrared measurements of CuInP$_2$S$_6$ and for reliably identifying its competing structural, ferroelectric, and antiferroelectric phases across temperature- and pressure-driven phase transitions.

\end{abstract}

\maketitle

%\tableofcontents

\section{\label{sec:introduction}Introduction}

CuInP$_2$S$_6$ (CIPS) is a layered van der Waals ferroelectric material that has attracted significant attention in recent years due to its interesting physical properties and promising technological applications~\cite{brehm2020tunable, mikolajick2021next, liu2016room, zhang2023ferroelectric, scott1989ferroelectric, zhang2023origin, dai2019two,    sun2025sliding}. In CIPS, the Cu ions play a crucial role in determining the ferroelectric behavior, as their displacement is primarily responsible for the spontaneous polarization of the crystal. This property makes  CIPS particularly attractive for a range of emerging technologies, including next-generation data storage devices such as nonvolatile FeRAM memories~\cite{mikolajick2021next, scott1989ferroelectric, zhang2023origin}, flexible electronics~\cite{dai2019two}, and advanced optoelectronic applications~\cite{sun2025sliding, studenyak2003optical}.

\textcolor{black}{CIPS features distinctive structural characteristics that give rise} to several remarkable properties, including negative piezoelectricity~\cite{you2019origin, neumayer2019giant}, quadruple- and even sextuple-well ferroelectric energy landscapes~\cite{neumayer2025competing, yu2023polarization, li2024realization, brehm2020tunable}, which originate from Cu-ion displacements both within the CIPS monolayer and across the van der Waals gap~\cite{seleznev2023cyclic,neumayer2020alignment,mohammadi2025pressure, zhang2021anisotropic, liang2025configurable, neal2022vibrational, zhou2020anomalous, morozovska2023anomalous}. In addition, CIPS exhibits high ionic conductivity~\cite{liu2016room, zhang2021anisotropic, maisonneuve1997ionic, simon1994paraelectric, bourdon199931p, neumayer2022ionic, sun2019strain}, ultrafast polarization switching near room temperature~\cite{liu2016room, neumayer2020alignment, zhou2020anomalous, jiang2024dual, morozovska2024ferri, he2023unconventional}, and quantized adiabatic charge transfer~\cite{seleznev2023cyclic}, making it a versatile platform for studying coupled ionic and ferroelectric phenomena. Importantly, these properties are strongly governed by the crystal symmetry and phase of the material.

A fundamental characteristic of ferroelectric materials is the Curie temperature ($T_c$), below which the structure undergoes a phase transition from a paraelectric to a ferroelectric state. In CIPS, the Curie temperature is close to room temperature ($T_c \sim 310$~K)~\cite{grzechnik1998pressure, maisonneuve1995room, fagot2003interplay, dziaugys2020piezoelectric}. 
Below $T_c$, the CIPS crystal transforms from the paraelectric $C2/c$ phase to the ferroelectric $Cc$ phase, primarily driven by the off-centering instability of the Cu ions.
Beyond this temperature-driven transition, CIPS exhibits a remarkably rich structural phase diagram under applied pressure, with a multitude of pressure-induced phases reported in the literature~\cite{shah2026pressure, mohammadi2025pressure, li2026ferroelectricity}.
Definitive experimental identification of these phases, however, remains challenging using conventional x-ray diffraction techniques, particularly when the competing phases exhibit subtle structural differences~\cite{shah2026pressure, bu2022enhanced, zhou2024sliding, yao2023anomalous, fang2023significant}.

Vibrational spectroscopy is a sensitive probe of crystal symmetry and local chemical bonding, making it particularly valuable for distinguishing the  competing structural phases of CIPS.
Despite extensive experimental studies~\cite{zhou2024sliding, neumayer2025competing, bu2022enhanced, liang2025configurable}, definitive vibrational fingerprints for its various temperature- and pressure-induced phases remain poorly established.
This challenge is further complicated in Raman measurements by the strong dependence of the observed modes and their intensities on crystal orientation, light polarization, and Raman-scattering geometry~\cite{neal2022vibrational, zhou2023unconventional, grzechnik1998pressure}. 
Consequently, different experimental configurations can probe different subsets of symmetry-allowed Raman modes, making direct comparison among reported results difficult~\cite{yao2023anomalous, li2024realization, rao2023raman}.
Hence, a comprehensive theoretical framework that establishes phase-specific vibrational fingerprints together with polarization- and geometry-dependent Raman selection rules is  essential for unambiguous phase determination and consistent interpretation of Raman measurements in CIPS.

%These considerations highlight the importance of accurate structural phase identification for understanding the phase transitions and the corresponding physical properties of CIPS. Accordingly, the pressure-induced structural evolution of CIPS has been extensively investigated using various experimental techniques, including X-ray diffraction (XRD)~\cite{bu2022enhanced, zhou2024sliding, yao2023anomalous, fang2023significant}, transmission electron microscopy (TEM)~\cite{susner2017metal, liu2016room}, and Raman and infrared (IR) spectroscopy~\cite{guranich2007influence, neal2022vibrational, zhou2023unconventional, rao2021pressure, sun2025sliding, grzechnik1998pressure, rao2023raman, fang2023significant}, among others.

Motivated by the rich pressure- and temperature-dependent behavior of CIPS, numerous studies have focused on identifying its structural phases and mapping the corresponding phase transitions. Bu \textit{et al.}~\cite{bu2022enhanced} reported, through in situ Raman spectroscopy and x-ray diffraction (XRD) measurements under varying pressures, a phase transition from the monoclinic ferroelectric $Cc$ phase to a paraelectric trigonal $P\overline{3}1m$ phase. Yao \textit{et al.}~\cite{yao2023anomalous}, employing single-crystal X-ray diffraction (SCXRD) and in situ Raman spectroscopy, investigated the evolution of Cu occupancy and migration with increasing pressure and correlated these changes with studying their crystal structure. Fang \textit{et al.}~\cite{fang2023significant} identified a four-stage evolution of photoresponsivity under compression, which they associated with the $Cc$ to $P\overline{3}1m$ phase transition.

Subsequently, Rao \textit{et al.}~\cite{rao2021pressure} analyzed pressure-dependent Raman peak shifts and suggested that the monoclinic-to-trigonal phase transition occurs near 5~GPa. Zhou \textit{et al.}~\cite{zhou2024sliding}, combining SCXRD and Raman spectroscopy, constructed a pressure--temperature phase diagram of CIPS and further clarified the monoclinic-to-trigonal transition. More recently, Shah \textit{et al.}~\cite{shah2026pressure} employed XRD together with Raman and IR spectroscopy to examine the pressure-induced phase sequence of CIPS by following the vibrational modes and the disappearance of characteristic diffraction peaks. Their results revealed a pressure-induced  structural phase transition from the monoclinic $Cc$ phase to the trigonal $P31c$ phase. 

%Although these previous studies have utilized diverse experimental techniques to identify structural phases in van der Waals CIPS, unambiguous phase determination based solely on XRD can be challenging; thus, vibrational spectroscopy offers a useful alternative. In particular, assigning specific symmetry phases through Raman and IR mode analysis requires careful consideration of polarization selection rules and mode symmetries.

In this work, we employ a first-principles theoretical approach to systematically investigate all the reported monoclinic and trigonal crystallographic phases of CIPS, including the ferroelectric $Cc$ and $P31c$, the paraelectric $C2/c$ and $P\overline{3}1c$, and the antiferroelectric $P\overline{1}$ and $P\overline{3}$ phases that emerge under different pressure and temperature conditions. By calculating the phonon spectra, we identify distinct vibrational signatures that allow differentiation between monoclinic and trigonal symmetries, as well as their polar and nonpolar states. Furthermore, through systematic analysis of Raman tensors and polarization selection rules within the framework of Porto notation, we determine the active and silent Raman modes under various experimental scattering geometries. This combined symmetry-resolved vibrational and polarization analysis provides a comprehensive fingerprint for reliable phase identification of CIPS.

\section{\label{sec:methods}Methods}

All density functional theory (DFT) calculations were performed using the Vienna \textit{ab initio} Simulation Package (VASP)~\cite{KRESSE199615,PhysRevB.54.11169,PhysRevB.59.1758} within the projector augmented-wave (PAW) framework. 
The generalized gradient approximation (GGA) parametrized by Perdew, Burke, and Ernzerhof for solids (PBEsol)~\cite{PhysRevLett.100.136406} was employed. A kinetic-energy cutoff of 400~eV was used for the plane-wave basis set. The electronic self-consistency criterion was set to $10^{-7}$~eV, and the force-convergence threshold for the relaxation of atomic coordinates was set to $10^{-3}$~eV\,\AA$^{-1}$.The DFT-D3 method was incorporated to account for weak vdW interactions~\cite{https://doi.org/10.1002/jcc.21759,10.1063/1.3382344}.  The reciprocal space of all phases was sampled using a $\Gamma$-centered $8 \times 8 \times 4$ \textit{k}-mesh.

Phonon dispersions, shown in the Supplemental Materials (SM)~\cite{SM}, for all phases were computed using $2 \times 2 \times 1$ supercells. The non-analytical term correction (NAC) was applied using the Gonze scheme~\cite{PhysRevB.55.10355, PhysRevB.50.13035} as implemented in the \texttt{phonopy} package~\cite{togo_first_2015}. The zone-center phonon eigenvectors were then used to simulate the theoretical infrared and Raman spectra following the methodology described in Ref.~\cite{skelton_lattice_2017}. The Bilbao Crystallographic Server~\cite{Kroumova01012003} and the \texttt{phonopy} package~\cite{togo_first_2015} were employed to determine the Raman selection rules and the symmetry characteristics of the phonon modes.

\begin{figure*}[t]
\centering
\includegraphics[width=\textwidth]{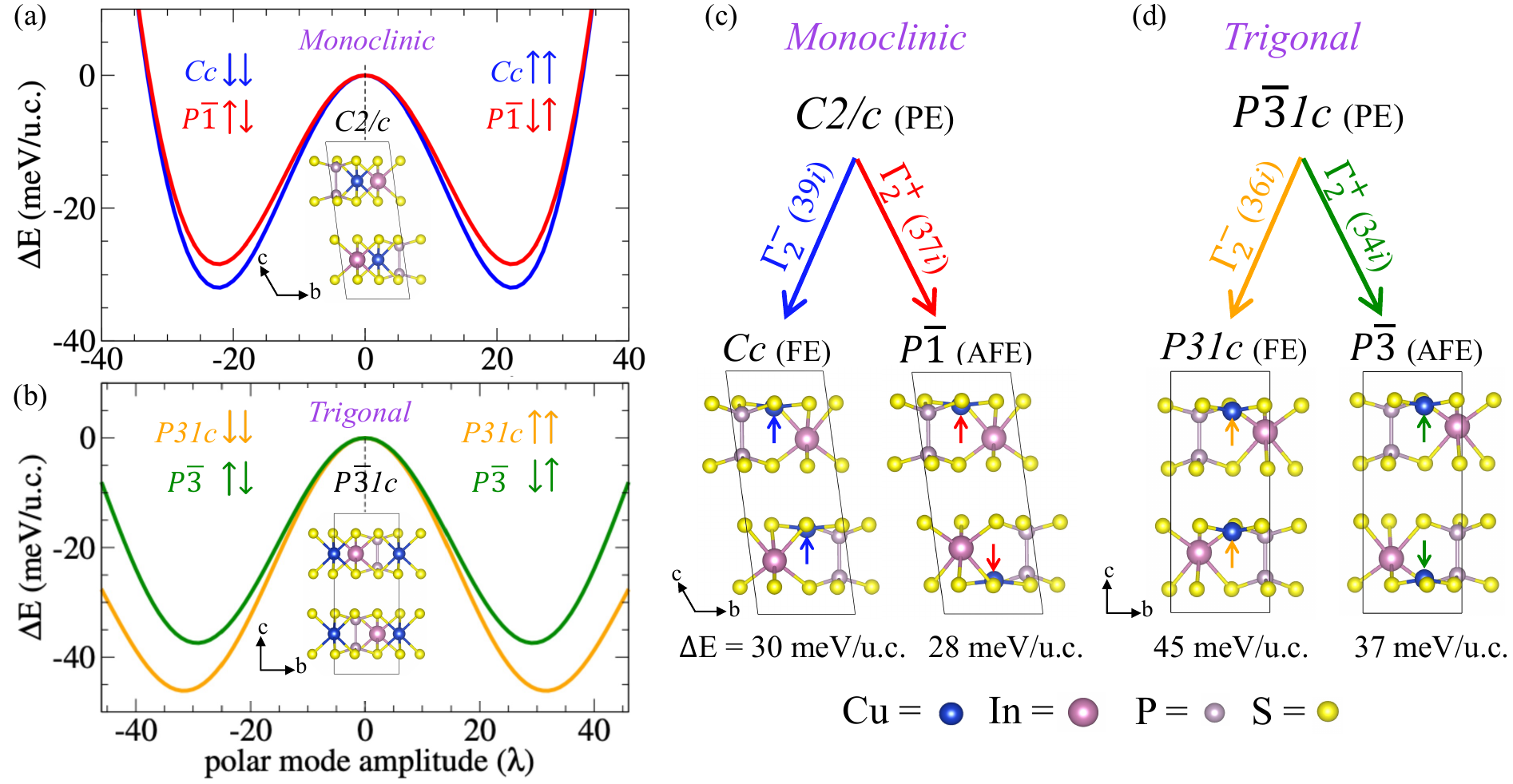}
\caption{\label{fig:modulation}
Double-well potential-energy profiles and structural evolution obtained by freezing the unstable polar $\Gamma_2^{-}$ and antipolar $\Gamma_2^{+}$ phonon modes of the reference paraelectric (PE) phases as a function of the phonon distortion amplitude. \textcolor{black}{The polar and antipolar distortions primarily involve out-of-plane Cu displacements relative to the surrounding S atoms, with parallel and oppositely directed Cu displacements, respectively.}
(a) For the monoclinic $C2/c$ phase, the $\Gamma_2^{-}$ and $\Gamma_2^{+}$ modes lead to the local minima corresponding to the ferroelectric (FE) $Cc$ and antiferroelectric (AFE) $P\overline{1}$ phases, respectively.
(b) Similarly, for the trigonal $P\overline{3}1c$ phase, these unstable modes lead to the FE $P31c$ and AFE $P\overline{3}$ phases, respectively.
\textcolor{black}{(c) and (d) illustrate the structural evolution from the monoclinic $C2/c$ and trigonal $P\overline{3}1c$ reference phases, respectively, along the polar and antipolar distortion pathways.}
The values in parentheses indicate the magnitudes of the imaginary phonon frequencies (cm$^{-1}$) (see Fig.~S1 in SM~\cite{SM}) of the corresponding unstable modes in the PE reference structures.
The energy differences of the FE and AFE phases are given relative to their corresponding PE reference phases, with total energies of $-100.57$~eV/u.c. for $C2/c$ and $-100.60$~eV/u.c. for $P\overline{3}1c$.
}
\end{figure*}

\section{\label{sec:Result}Results and Discussion}

\subsection{Analyzing the symmetries and properties of CuInP$_2$S$_6$}

\subsubsection{Energy Landscape and Ferroelectricity}

In the van der Waals layered bulk CIPS material, the Cu atom plays a key role in defining the net polarization. By moving the Cu atom from the centrosymmetric position toward the out-of-plane direction, inversion symmetry is broken, resulting in a non-zero ferroelectric polarization. In CIPS, if the Cu$^{+}$ atoms in two adjacent monolayers are located on opposite sides of their respective S  planes, each monolayer develops a finite polarization; however, because these polarizations have equal magnitude but opposite directions ($\uparrow\downarrow$ or $\downarrow\uparrow$), the net polarization of the structure becomes zero, corresponding to an antiferroelectric phase, as shown in Fig.~\ref{fig:modulation}. 
In contrast, when the Cu$^{+}$ atoms in adjacent monolayers are displaced toward the same side of the S plane, their polarizations align in the same direction  ($\uparrow\uparrow$ or $\downarrow\downarrow$), resulting in a non-zero ferroelectric polarization.

In CIPS, the paraelectric phases (shown in Fig.~\ref{fig:modulation}), corresponding to the monoclinic $C2/c$ (no.~15) and trigonal $P\overline{3}1c$ (no.~163) space groups, are stabilized above their respective $T_c$. \textcolor{black}{The Curie temperature of CIPS is strongly pressure dependent, decreasing with increasing pressure~\cite{10.1063/5.0299899}.} At temperatures below the corresponding transition temperatures, Cu off-centering from the centrosymmetric reference structures along unstable phonon modes leads to symmetry breaking and the stabilization of the polar ($Cc$ and $P31c$) phases. The corresponding antipolar distortions lead to the nonpolar ($P\overline{1}$ and $P\overline{3}$) phases, as illustrated in Fig.~\ref{fig:modulation}.

In addition to temperature, external pressure also plays an important role in tuning structural stability; under hydrostatic pressure at temperatures below $T_c$, the monoclinic $Cc$ phase undergoes a transition to the trigonal $P31c$ phase, which becomes stable at elevated pressures~\cite{shah2026pressure}.

In monoclinic  $C2/c$ phase, DFT calculations at 0\,K reveal two unstable phonon modes at Brillouin zone center, which are denoted as $\Gamma_2^{+}$ and $\Gamma_2^{-}$.
From a group-theoretical perspective, these modes correspond to $B_g$ and $B_u$ symmetries, respectively. 
These modes primarily represent an out-of-plane displacement of the two Cu ions located in adjacent CIPS layers. 
The $\Gamma_2^{-}$ mode corresponds to the case in which the Cu atoms displace from the midplane of the CIPS monolayer in the same direction toward the S octahedral plane ($\uparrow\uparrow$ or $\downarrow\downarrow$), leading to the ferroelectric $Cc$ (no.~9) phase. In contrast, the $\Gamma_2^{+}$ mode corresponds to the out-of-phase displacement of two Cu atoms from the midplane toward opposite sides of the S planes ($\uparrow\downarrow$ or $\downarrow\uparrow$), resulting in an antiferroelectric $P\overline{1}$ (no.~2) phase, as illustrated by the blue and red arrows in Fig.~\ref{fig:modulation}(a), respectively.

A similar scenario is observed for the trigonal $P\overline{3}1c$ phase, where two unstable modes, $\Gamma_2^{+}$ and $\Gamma_2^{-}$, lead to an antiferroelectric $P\overline{3}$ phase (no.~147) 
and a ferroelectric $P31c$ phase (no.~159), respectively, as depicted in Fig.~\ref{fig:modulation}(b).
Interestingly, these two unstable modes are responsible for an asynchronous ferroelectric switching mechanism in CIPS~\cite{mohammadi2025pressure}.

%In the trigonal $P\overline{3}1c$ phase, a similar scenario is observed. The $P\overline{3}1c$ phase exhibits two unstable phonon modes, denoted as $\Gamma_2^{+}$ and $\Gamma_2^{-}$, corresponding to $A_g$ and $A_u$ symmetries, respectively. Here, the $\Gamma_2^{-}$ mode corresponds to the in-phase vertical displacement of both Cu atoms ($\uparrow\uparrow$ or $\downarrow\downarrow$) yielding a ferroelectric $P31c$ phase (no.~159). Whereas, the $\Gamma_2^{+}$ mode corresponds to the out-of-phase vertical displacement of both Cu atoms ($\uparrow\downarrow$ or $\downarrow\uparrow$) resulting in an antiferroelectric $P\overline{3}$ phase (no.~147), as illustrated by the orange and green arrows in Fig.~\ref{fig:modulation}(b). 

%In the $\Gamma_2^{-}$ mode, the Cu atoms displace from the midplane of the CIPS monolayer in the same direction in adjacent layers ($\uparrow\uparrow$ or $\downarrow\downarrow$), leading to a phase transition from the paraelectric $P\overline{3}1c$ phase to the ferroelectric $P31c$ phase (No.~159). In contrast, following the $\Gamma_2^{+}$ mode, the Cu atoms displace in opposite directions in adjacent CIPS monolayers ($\uparrow\downarrow$ or $\downarrow\uparrow$), resulting in a phase transition to the antiferroelectric $P\overline{3}$ phase (No.~147), as illustrated by the orange and green arrows in Fig.~\ref{fig:modulation}b, respectively. This behavior has been investigated in detail in Ref.~\cite{mohammadi2025pressure}.

\begin{figure*}[htbp!]
\centering
\includegraphics[width=\textwidth]{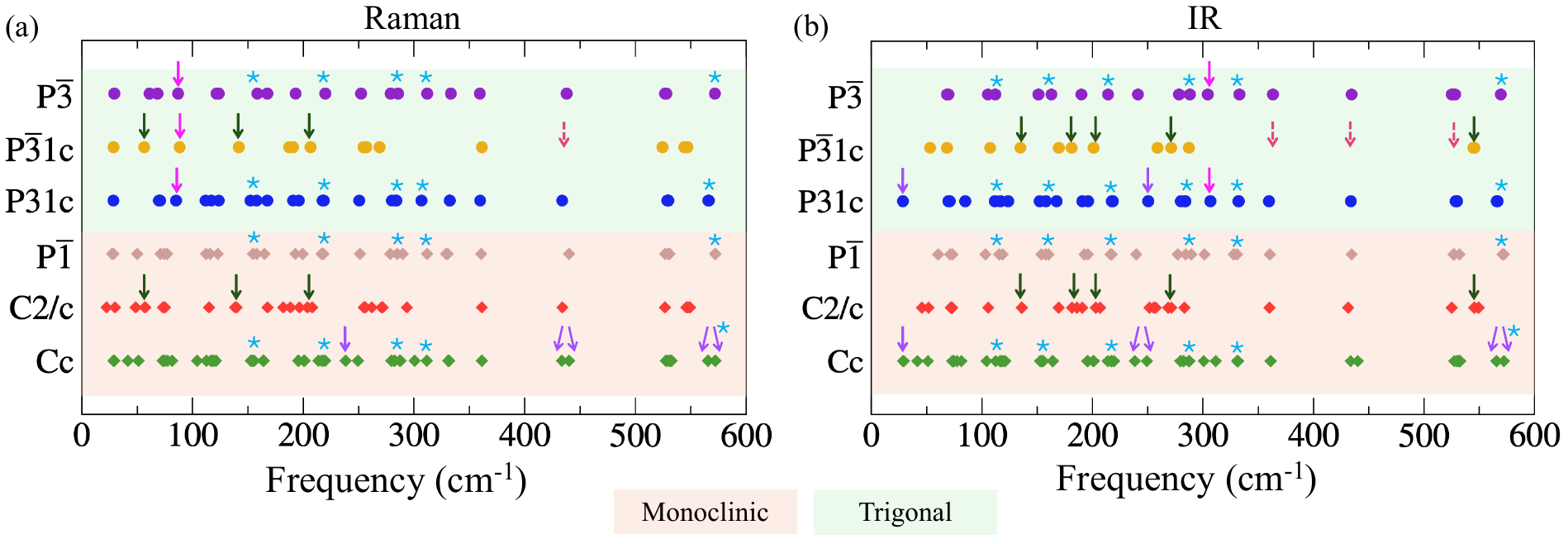}
\caption{\label{fig:Raman_IR}
Calculated (a) Raman and (b) infrared (IR) peak positions for the ferroelectric, paraelectric, and antiferroelectric phases of CuInP$_2$S$_6$. The light-red and light-green shaded regions distinguish the monoclinic and trigonal phases, respectively. Dark-green arrows (\textcolor{darkgreen}{$\boldsymbol{\downarrow}$}) indicate characteristic modes of the paraelectric phases, while pink arrows (\textcolor{magenta}{$\boldsymbol{\downarrow}$}) identify characteristic modes of the trigonal phases. Cyan stars (\textcolor{cyan}{$\boldsymbol{\ast}$}) denote modes common to the ferroelectric and antiferroelectric phases. Purple arrows (\textcolor{darkpurple}{$\boldsymbol{\downarrow}$}) highlight characteristic modes of the stable ferroelectric $Cc$ and $P31c$ phases, while dashed arrows indicate characteristic features (absence of modes) associated with the $P\overline{3}1c$ phase. 
One should particularly note the splitting of some peaks as system transitions from the high-symmetry trigonal to the low-symmetry monoclinic phase. 
}
\end{figure*}

The calculated energy profiles shown in Fig.~\ref{fig:modulation} are obtained as a function of the phonon distortion amplitude. We find that in both the $C2/c$ and $P\overline{3}1c$ phases, in response to the polar distortion of the Cu ions ($\sim 0.02$~\AA), the In ions exhibit a relatively small polar distortion ($\sim 0.002$~\AA), which is approximately one order of magnitude smaller than that of the Cu ions and therefore can be considered negligible. In Fig.~\ref{fig:modulation}, we start from the high-energy and dynamically unstable $C2/c$ and $P\overline{3}1c$ phases, and by following their corresponding unstable polar modes, we reach the ferroelectric $Cc$ and antiferroelectric $P\overline{1}$ phases from the monoclinic $C2/c$ phase, and the ferroelectric $P31c$ and antiferroelectric $P\overline{3}$ phases from the trigonal $P\overline{3}1c$ phase. \textcolor{black}{These antiferroelectric configurations can serve as metastable intermediate states during electric-field-driven polarization switching, as discussed in Ref.~\cite{mohammadi2025pressure}.}The optimized lattice parameters for all candidate structures are summarized in Table~S1 of the Supplementary Data \cite{SM}.

Based on recent studies on CIPS under hydrostatic pressure, the pressure-induced phase transitions have been investigated both experimentally, using Raman and infrared spectroscopy~\cite{grzechnik1998pressure, rao2023raman, shah2026pressure}, and theoretically, through enthalpy calculations~\cite{mohammadi2025pressure, zhou2024sliding}. These studies show that by applying hydrostatic pressure at temperatures below $T_c$, a phase transition occurs from the ferroelectric $Cc$ phase to the ferroelectric $P31c$ phase, and at high pressures, the system remains stable in the $P31c$ phase. Therefore, depending on the applied pressure range, two stable ferroelectric phases, $Cc$ and $P31c$, can be realized in CuInP$_2$S$_6$.
However, ultrahigh pressures are expected to suppress any out-of-plane ferroelectric distortions and stabilize the paraelectric $P\overline{3}1c$ phase.

\subsubsection{Selection Rules in All CIPS Candidate Structures}

Having identified the paraelectric ($C2/c$, $P\overline{3}1c$), ferroelectric ($Cc$, $P31c$), and antiferroelectric ($P\overline{1}$, $P\overline{3}$) phases (see Fig.~\ref{fig:modulation}), it is essential to further examine their dynamical stability, including the LO--TO corrections ~\cite{PhysRevB.55.10355, PhysRevB.50.13035} across the Brillouin zone, as shown in Fig.\,S1~\cite{SM}. The results demonstrate that the ferroelectric and antiferroelectric phases, which emerge from the double-well energy landscape of their corresponding paraelectric phases, are dynamically stable and exhibit no imaginary phonon frequencies throughout the Brillouin zone. In contrast, the paraelectric phases display unstable phonon modes, consistent with the discussion above and previous studies~\cite{mohammadi2025pressure}.

In all considered CIPS crystal structures, the primitive cell contains two formula units, giving 20 atoms per cell. Therefore, 60 phonon modes are expected in total. Among these, 3 are acoustic modes, while the remaining 57 are optical modes at the Brillouin zone center. According to group theory, these zone-center vibrational modes can be decomposed into irreducible representations as given below for the two ferroelectric phases, $Cc$ and $P31c$, which are stable phases of CIPS. The corresponding irreducible representations for the paraelectric and antiferroelectric phases of the monoclinic and trigonal structures are provided in the Section S3 of SM~\cite{SM}.

 For the ferroelectric monoclinic $Cc$ phase:
\begin{equation}
\begin{aligned}
\Gamma_{\text{acoustic}} &= 2A' \oplus A'', \\
\Gamma_{\text{optical}}  &= 28A' \oplus 29A''.
\end{aligned}
\label{eq:Cc_modes}
\end{equation}

For the ferroelectric trigonal $P31c$ phase:
\begin{equation}
\begin{aligned}
\Gamma_{\text{acoustic}} &= A_1 \oplus E, \\
\Gamma_{\text{optical}}  &= 9A_1 \oplus 10A_2 \oplus 19E.
\end{aligned}
\label{eq:P31c_modes}
\end{equation}

Table~\ref{tab:Raman/IR} summarizes the Raman- and infrared-active modes for the six considered phases, while Fig.~\ref{fig:Raman_IR} presents the corresponding DFT calculated peak positions for all ferroelectric, paraelectric, and antiferroelectric candidates in both trigonal and monoclinic crystal structures of CIPS. The complete set of DFT calculated Raman- and infrared-active mode frequencies for all phases is provided in Table~S2 and Table~S3 of SM~\cite{SM}, respectively \textcolor{black}{(also see Fig.~S3 of the SM~\cite{SM} for the corresponding calculated Raman and IR intensities)}.
Several distinct spectral features are identified, which can facilitate reliable and systematic phase identification of the CIPS structures using vibrational spectroscopy.

\begin{table}[htbp!]
\centering
\renewcommand{\arraystretch}{1.3}
\setlength{\tabcolsep}{6pt}
\caption{Raman- and infrared-active mode symmetries in the studied phases of CuInP$_2$S$_6$.}
\label{tab:Raman/IR}
\begin{tabular}{l l c c}
\hline\hline
\textbf{Type} & \textbf{Phase} & \textbf{Raman} & \textbf{IR} \\
\hline

\multirow{2}{*}{Ferroelectric}
& $Cc$      & $A',\, A''$      & $A',\, A''$ \\
& $P31c$   & $A_1,\, E$       & $A_1,\, E$ \\

\hline

\multirow{2}{*}{Paraelectric}
& $C2/c$            & $A_g,\, B_g$     & $A_u,\, B_u$ \\
& $P\overline{3}1c$ & $A_{1g},\, E_g$  & $A_{2u},\, E_u$ \\

\hline

\multirow{2}{*}{Antiferroelectric}
& $P\overline{1}$   & $A_g$            & $A_u$ \\
& $P\overline{3}$   & $A_g,\, E_g$     & $A_u,\, E_u$ \\

\hline\hline
\end{tabular}
\end{table}
In the Raman spectra [Fig.~\ref{fig:Raman_IR}(a)], characteristic signature modes allow clear distinction between different crystal symmetries and phases. In the $Cc$ phase, most peaks arise from nearly degenerate $A'$ and $A''$ modes, as the frequency difference between these modes is negligible due to the weak coupling between adjacent CIPS layers. A key feature of all trigonal phases is the presence of a mode near  86~cm$^{-1}$ with E and $E_g$ symmetries, which is absent in monoclinic phases and thus serves as a clear symmetry indicator. The paraelectric phases are characterized by modes near 56,  140, and  207~cm$^{-1}$, associated with $B_g$ and $E_g$ symmetries. In contrast, modes appearing at  157,  217,  284,  331, and  570~cm$^{-1}$ are exclusive to the ferroelectric and antiferroelectric phases and are associated with Cu displacements toward the S octahedral planes. The trigonal paraelectric $P\overline{3}1c$ phase can be further identified by the absence of a mode near  433~cm$^{-1}$. Finally, the $Cc$ phase is characterized by the presence of a distinct peak near  238~cm$^{-1}$, which is absent in all other phases, and by a characteristic splitting of a high-frequency mode into two peaks near 565 and 572~cm$^{-1}$, whereas this feature appears as a single peak in the other phases. \textcolor{black}{The atomic displacement patterns associated with the characteristic Raman modes of the $Cc$ phase and the prominent modes of the trigonal phases are illustrated in Fig.~S4 of the SM.\cite{SM}}

The infrared spectra [Fig.~\ref{fig:Raman_IR}(b)] provide complementary signatures. The paraelectric phases show distinct modes near 135,  181,  202, 271, and  545~cm$^{-1}$, corresponding to $A_u$ and $E_u$ symmetries, which are absent in other phases. In contrast, the ferroelectric and antiferroelectric phases exhibit modes near  115,  152,  216, 279, 331, and  566~cm$^{-1}$. A distinguishing feature of trigonal FE and AFE phases is the presence of a mode near  305~cm$^{-1}$, associated with $A_u$ and $A_1$ symmetries, which is absent in the monoclinic counterparts. The $P\overline{3}1c$ phase is further characterized by the absence of modes near  363, 434, and  527~cm$^{-1}$. To distinguish between the two ferroelectric phases, a low-frequency mode near  28~cm$^{-1}$ is present, while the $P31c$ phase exhibits modes near 250 and 566~cm$^{-1}$ that split into pairs at  238/ 249 and  565/ 572~cm$^{-1}$ in the $Cc$ phase. \textcolor{black}{It is worth noting that the unstable $B_g$ and $B_u$ modes of the monoclinic $C2/c$ phase are Raman- and IR-active, respectively, while the unstable $A_{2u}$ mode of the trigonal $P\overline{3}1c$ phase is IR-active. Although these modes are dynamically unstable at 0~K, finite-temperature effects and, in the trigonal phase, applied pressure may stabilize the corresponding lattice vibrations.}

Overall, these distinct Raman and infrared spectral features provide definitive vibrational fingerprints for identifying both the crystal symmetry and the stable phases of CIPS.

\subsection{Porto Notation and Raman Selection Rules}

To analyze the polarization- and sample-orientation-dependent Raman response of CIPS and to compare our results with experimental Raman scattering measurements, we analyze Porto notation~\cite{porto1967raman, damen1966raman}. This notation describes the orientation of the crystal with respect to the polarization directions of the incident and scattered light in a Raman experiment. The Porto notation for Raman scattering is written as

\begin{equation}
k_i \, \left(\hat{\mathbf{e}}_i\, \hat{\mathbf{e}}_s\right) \, k_s,
\end{equation}

where $k_i$ denotes the propagation direction of the incident light, $\hat{\mathbf{e}}_i$ is the polarization direction of the incident light, $\hat{\mathbf{e}}_s$ represents the polarization direction of the scattered light, and $k_s$ corresponds to the propagation direction of the scattered light.

\begin{figure}[htbp!]
\centering
\includegraphics[width=0.5\textwidth]{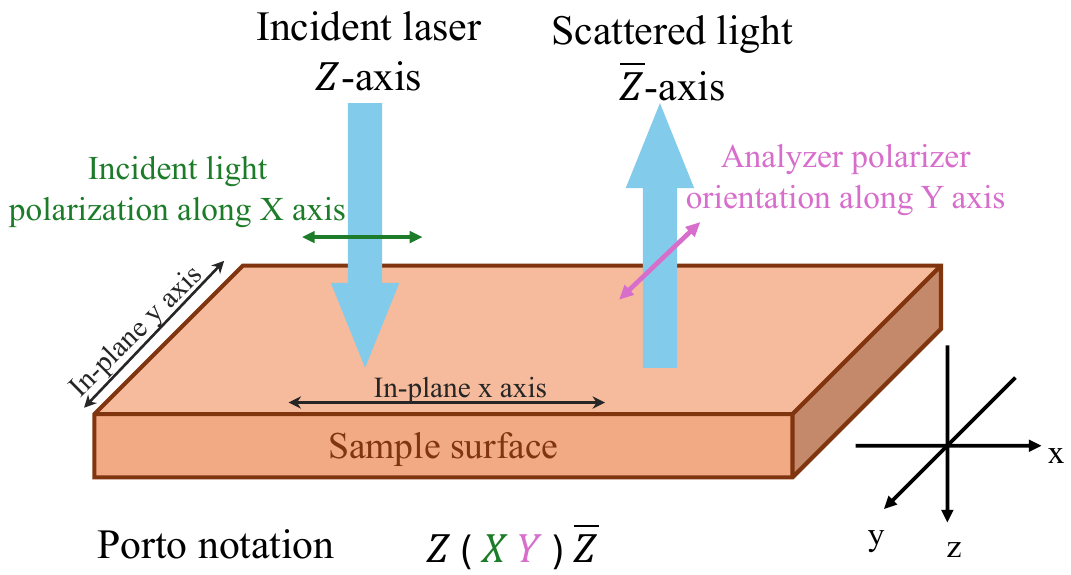} 
\caption{\label{fig:porto's}
Schematic illustration of the Raman scattering geometry described using the Porto notation. The incident (scattered) light propagates along the $Z$ ($\overline{Z}$) direction and is polarized along the in-plane orientation. \textcolor{black}{The coordinate axes represent the tensor-aligned scattering geometry used to determine the Raman selection rules. For the monoclinic case, unique axis $\hat{y} \parallel \mathbf{b}$, while for the trigonal case, $\hat{z} \parallel \mathbf{c}$.}}
%while the scattered light is collected along the $\overline{Z}$ direction with polarization analyzed along the $Y$ axis, corresponding to the $Z(XY)\overline{Z}$ configuration.}
\end{figure}

Figure~\ref{fig:porto's} illustrates the experimental geometry for Raman scattering. As an example, in the $Z(XY)\overline{Z}$ configuration, the incident light propagates along the $Z$ direction, while the scattered light is collected along the opposite direction, $\overline{Z}$. In this configuration, the polarization of the incident light is aligned along the $X$ direction, and the polarization of the scattered light is oriented along the $Y$ direction.

\begin{table*}[htbp!]
\centering
\caption{Raman-active modes under different Raman backscattering geometries for the studied phases of layered CuInP$_2$S$_6$.
}
\label{tab:raman_geometry}
\renewcommand{\arraystretch}{1.2}
\setlength{\tabcolsep}{7pt}  
\begin{tabular}{c c c c c c c}
\hline\hline
\diagbox[width=9em]{Geometry}{Phase}
& \textbf{Cc}
& \textbf{C2/c}
& $\boldsymbol{\mathrm{P\bar{1}}}$
& \textbf{P31c}
& $\boldsymbol{\mathrm{P\bar{3}1c}}$
& $\boldsymbol{\mathrm{P\bar{3}}}$ \\
\hline

$Z(XX)\bar{Z}$ 
& $A'(\mathrm{TO})$ 
& $A_g$ 
& $A_g$ 
& $A_1(\mathrm{LO}),\,E(\mathrm{TO})$ 
& $A_{1g},\,E_g$ 
& $^{1}E_g,\,^{2}E_g,\,A_g$ \\

$Z(XY)\bar{Z}$ 
& \textcolor{black}{$A''(\mathrm{TO})$} 
& \textcolor{black}{$B_g$}
& $A_g$ 
& $E(\mathrm{TO})$ 
& $E_g$ 
& $^{1}E_g,\,^{2}E_g$ \\

$Z(YY)\bar{Z}$ 
& $A'(\mathrm{TO})$ 
& $A_g$ 
& $A_g$ 
& $A_1(\mathrm{LO}),\,E(\mathrm{TO})$ 
& $A_{1g},\,E_g$ 
& $^{1}E_g,\,^{2}E_g,\,A_g$ \\

\hline\hline
\end{tabular}
\end{table*}

For the ferroelectric phases of CIPS, $Cc$ and $P31c$, the optical $A'$ and $A''$ phonon modes in the $Cc$ phase, as well as the $A_1$ and $E$ modes in the $P31c$ phase, are first-order Raman active. Since both phases are non-centrosymmetric, their Raman activity is governed directly by the symmetry-allowed components of the polarizability tensor and the scattering geometry.

\textcolor{black}{For the monoclinic ferroelectric $Cc$ phase (point group $C_s$), the mirror plane is perpendicular to the crystallographic $\mathbf{b}$ axis in the conventional unique-$\mathbf{b}$ setting. We therefore adopt an orthonormal Cartesian coordinate system with $\hat{y}\parallel\mathbf{b}$, such that the $xz$ plane coincides with the crystallographic mirror plane.
For a (001)-oriented crystal, we further choose $\hat{z}\parallel\mathbf{c}^{*}$, normal to the layers, and $\hat{x}=\hat{y}\times\hat{z}$ along the remaining in-plane direction.
Under the mirror operation $M_y$,
the Raman tensors $\mathcal{R}(A')$ and $\mathcal{R}(A'')$
transform as even and odd representations, respectively:
$M_y\,\mathcal{R}(A')\,M_y^{T}=\mathcal{R}(A')$, and
$M_y\,\mathcal{R}(A'')\,M_y^{T}=-\mathcal{R}(A'')$.
Imposing these symmetry constraints on the Raman polarizability tensor yields the
following symmetry-allowed forms~\cite{porto1967raman, damen1966raman, didomenico1968raman}:
}
\begin{equation}
\mathcal{R}[A' (x,z)] =
\begin{pmatrix}
a & 0 & d \\
0 & b & 0 \\
d & 0 & c
\end{pmatrix},~~\text{and}
\end{equation}

\begin{equation}
\mathcal{R}[A''(y)] =
\begin{pmatrix}
0 & e & 0 \\
e & 0 & f \\
0 & f & 0
\end{pmatrix}.
\end{equation}

%\textcolor{black}{For the Raman polarization selection rules, the laboratory axes $(X,Y,Z)$ in Fig.~\ref{fig:porto's} are chosen to coincide with the Cartesian axes $(x,y,z)$ used to express the Raman tensors, such that $X\parallel x$, $Y\parallel y$, and $Z\parallel z$.}
%The Raman tensors $\alpha$ associated with each irreducible representation are given below~\cite{porto1967raman, damen1966raman, didomenico1968raman}:

\textcolor{black}{For the trigonal ferroelectric $P31c$ phase (point group $C_{3v}$), the threefold rotation axis is parallel to the crystallographic $\mathbf{c}$ axis, while the three vertical mirror planes contain this axis. We therefore adopt an orthonormal Cartesian coordinate system with $\hat{z}\parallel\mathbf{c}$, 
$\hat{y}$ lying within one of the vertical mirror planes, and $\hat{x}$ perpendicular to the $yz$ mirror plane.}
\textcolor{black}{The Raman-active modes belong to the $A_1$ and doubly degenerate $E$ irreducible representations.
Imposing the corresponding threefold rotational and mirror-symmetry constraints yields the following symmetry-allowed Raman tensors:
}

\begin{equation}
\mathcal{R}[A_1(z)]=
\begin{pmatrix}
a & 0 & 0\\
0 & a & 0\\
0 & 0 & b
\end{pmatrix},
\end{equation}
\begin{equation}
\mathcal{R}[E(x)] =
\begin{pmatrix}
0 & c & d \\
c & 0 & 0 \\
d & 0 & 0
\end{pmatrix},~\text{and}~
\mathcal{R}[E(y)] =
\begin{pmatrix}
c & 0 & 0 \\
0 & -c & d \\
0 & d & 0
\end{pmatrix}.
\end{equation}

Here, $x$, $y$, and $z$ denote the Cartesian polarization directions
of the incident and scattered light, while $a$, $b$, $c$, $d$, $e$, and $f$ represent the symmetry-allowed Raman tensor elements,
whose values depend on the specific phonon mode.

%The coordinates indicated in parentheses specify the direction of the phonon-induced polarizability variation; in particular, the $A'$ phonons are polarized within the $xy$ plane. In contrast, the $A''$ modes are polarized along the $z$ axis.

The scattering intensity of a first-order Stokes Raman transition~\cite{peter2010fundamentals,loudon2001raman} is given by
\begin{equation}
I \propto \left| \hat{\mathbf{e}}_s^{\mathrm T} 
~\mathcal{R}~  \hat{\mathbf{e}}_i \right|^2,
\end{equation}
where $\mathcal{R}$ is the Raman polarizability tensor associated with a given vibrational mode. Depending on the Raman scattering geometry and the polarization configuration selected in  experiments, the observed intensity may vary. If the specific Raman tensor element selected by the polarization configuration vanishes by symmetry, the intensity becomes zero ($I=0$), and the corresponding mode is Raman inactive (silent) in that geometry. Therefore, the appearance or suppression of specific phonon modes follows the Porto notation configuration.

\begin{figure}[htbp!]
\centering
\includegraphics[width=\columnwidth]{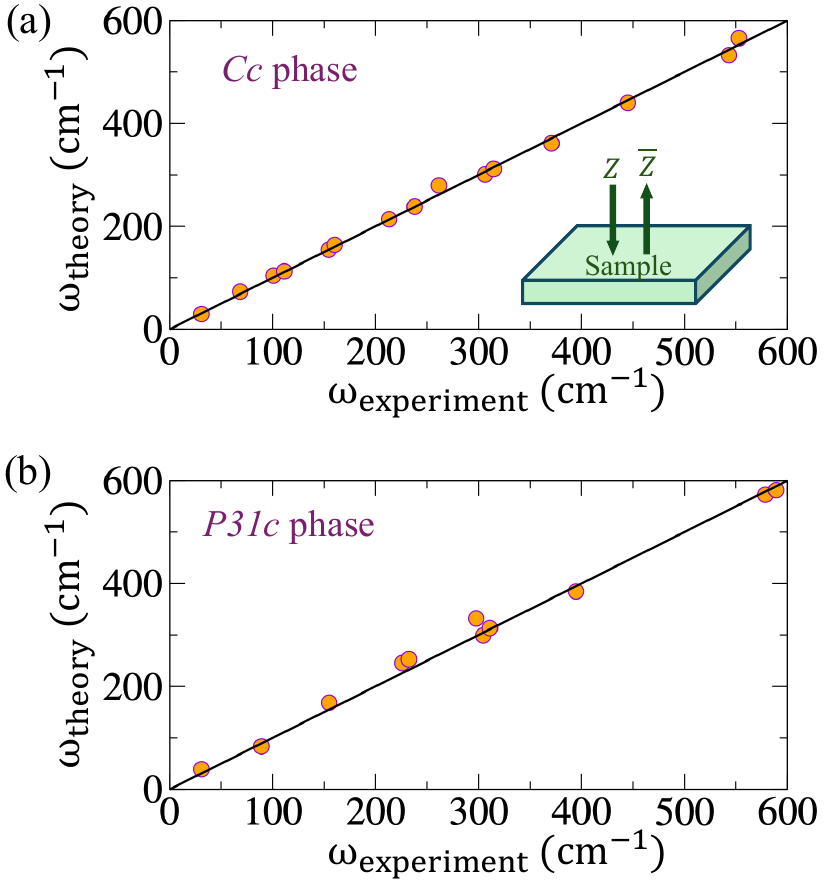}
\caption{\label{fig:exp_the}
Comparison between the DFT-calculated (PBEsol+D3) and experimental Raman mode frequencies for (a)\,low-pressure $Cc$ phase and (b)\,high-pressure $P31c$ phase. The inset illustrates the experimental backscattering geometry, in which the incident and scattered light propagate along the $Z$ and $\overline{Z}$ directions, respectively, normal to the sample surface, without a specific polarization selection. The experimental data are taken from Ref.~\cite{shah2026pressure}.}
\end{figure}

In polar insulators, long-range electrostatic forces associated with lattice ionicity lead to additional splitting of the vibrational modes. As a consequence, each polar optical mode splits into transverse optical (TO) and longitudinal optical (LO) components at non-zero wave vector (${\bf{q}} \rightarrow 0$). 
In our Raman geometry (Fig.~\ref{fig:porto's}), the LO modes correspond to vibrations polarized along the propagation direction of light (e.g., along the $Z$ direction in the schematic), whereas the TO modes are polarized within the plane perpendicular to the propagation direction (i.e., within the $xy$ plane).

Based on this, we summarize the Raman-active optical lattice vibrations in Table~\ref{tab:raman_geometry} for the configuration where the incident and scattered light propagate along the $Z$ direction and the sample surface lies in the $xy$ plane (which is a common geometry for layered CIPS materials), as illustrated in Fig.~\ref{fig:porto's}. Additional polarization configurations for all considered phases are provided in Table~S4 of SM~\cite{SM}. 

Since pressure-induced phase transitions in CIPS have been reported previously~\cite{mohammadi2025pressure, shah2026pressure}, we compare our calculated frequency of the Raman-active modes with the reported experimental Raman spectra measured under comparable scattering geometries~\cite{shah2026pressure}.
We obtain good agreement between the theoretical and experimental results, as shown in Fig.~\ref{fig:exp_the}, indicating that the experimental data at ambient pressure correspond to the $Cc$ phase, while those at higher pressures (around 8\,GPa) are consistent with the $P31c$ phase. \textcolor{black}{The theoretical Raman frequencies of the $P31c$ phase were also calculated at 8~GPa to account for the pressure-induced frequency shifts in the comparison. Although vibrational frequencies shift under pressure, the characteristic fingerprints remain identifiable, while symmetry-based Raman selection rules are preserved within each phase.}
This agreement further validates the PBEsol+D3 (DFT) level of theory for aptly describing the vibrational properties of CIPS and provides confidence in its use for identifying the vibrational fingerprints of competing phases. \textcolor{black}{The small differences between calculated and experimental frequencies may arise from anharmonic effects neglected in the harmonic approximation.}

\section{\label{sec:conclusion}Conclusion}

In summary, we have established a comprehensive first-principles framework for identifying the competing structural and polar phases of CuInP$_2$S$_6$ through their vibrational fingerprints. 
By systematically analyzing the phonon spectra of the paraelectric $C2/c$ and $P\overline{3}1c$, ferroelectric $Cc$ and $P31c$, and antiferroelectric $P\overline{1}$ and $P\overline{3}$ phases, 
we determine their Raman- and infrared-active modes and identify characteristic spectral features that can be used to distinguish among these closely related structures. 
Importantly, our analysis shows that phase identification based solely on the presence or absence of individual Raman peaks can be misleading without accounting for crystal orientation, polarization, and Raman-scattering geometry. Using symmetry analysis and Porto notation, we establish the corresponding Raman selection rules and determine which vibrational modes should be observable under different experimental configurations. Overall, this work provides a comprehensive vibrational fingerprint framework for distinguishing the structural phases of CIPS, which is crucial for accurate phase identification in experimental studies.

\section*{Acknowledgment}
Authors acknowledge support from the U.S.~Department of Energy, Office of Science, Office of Fusion Energy Sciences, Quantum Information Science program under Award No. DE-SC-0020340. Authors also thank the Pittsburgh Supercomputer Center (Bridges2) supported by the Advanced Cyberinfrastructure Coordination Ecosystem: Services \& Support (ACCESS) program, which is supported by National Science Foundation Grants No.~2138259, No.~2138286, No.~2138307, No.~2137603, and No.~2138296.
Authors also acknowledge support from the Furth Research Fund at the University of Rochester.

\bibliography{References}% Produces the bibliography via BibTeX.

\end{document}

\begin{center}
  {\LARGE \textbf{Supplementary Information}}\\[6pt]
  {\large for}\\[6pt]
  {\LARGE \textbf{Vibrational Fingerprints of Competing Structural and Polar Phases in CuInP$_2$S$_6$}}\\[12pt]
  \normalsize
  \textbf{Pegah Mohammadi}$^{a}$, \textbf{and Sobhit Singh}$^{a,b}$\\[4pt]
  \small
  $^{a}$Department of Mechanical Engineering, University of Rochester, Rochester, NY 14627, USA\\
  $^{b}$Materials Science Program, University of Rochester, Rochester, NY 14627, USA\\[2pt]
  Corresponding author: \href{mailto:s.singh@rochester.edu}{s.singh@rochester.edu}
\end{center}

\vspace{12pt}
\hrule
\vspace{6pt}
\noindent \textbf{Contents}\\
S1. Optimized Lattice Parameters \dotfill \pageref{sec:lattice}\\
S2. Phonon Calculations of All Phases \dotfill \pageref{sec:phonons}\\
S3. Irreducible Representations of Paraelectric and Antiferroelectric Phases \dotfill \pageref{sec:irreducible}\\
S4. Raman and IR Active Mode Frequencies for All CuInP$_2$S$_6$ Prototypes \dotfill \pageref{sec:Raman_IR}\\
S5. Raman and IR Intensities \dotfill \pageref{sec:Raman_IR_intensity}\\
S6. Characteristic Raman Mode Displacements \dotfill \pageref{sec:atomic_displacements}\\
S7. Polarization Selection Rules for the Studied Phases of CuInP$_2$S$_6$ \dotfill \pageref{sec:porto_notation}\\
\vspace{6pt}
\hrule
\vspace{18pt}

\section{Optimized Lattice Parameters}
\label{sec:lattice}

In this section, we present the DFT-optimized lattice parameters for all considered structural candidates of CuInP$_2$S$_6$, including both monoclinic and trigonal crystallographic phases. The investigated structures comprise the ferroelectric $Cc$ and $P31c$ phases, the paraelectric $C2/c$ and $P\overline{3}1c$ phases, and the antiferroelectric $P\overline{1}$ and $P\overline{3}$ phases. 
%As expected, the paraelectric $C2/c$ phase and its dynamically stable derivatives, ferroelectric $Cc$ and antiferroelectric $P\overline{1}$ phases, belong to the monoclinic crystal, whereas the paraelectric $P\overline{3}1c$ phase and its dynamically stabilized derivatives, ferroelectric $P31c$ and antiferroelectric $P\overline{3}$ phases, trigonal crystals.

\begin{table*}[htbp!]
\centering
\caption{DFT-D3 optimized lattice parameters of the studied CuInP$_2$S$_6$ phases.
}
\label{tab:lattice_CIPS}
\renewcommand{\arraystretch}{1.25}
\setlength{\tabcolsep}{7pt}
\begin{tabular}{c c c c c c c c}
\hline\hline
\textbf{Type} & \textbf{Phase} 
& $a$ (\AA) & $b$ (\AA) & $c$ (\AA)
& $\alpha$ ($^\circ$) & $\beta$ ($^\circ$) & $\gamma$ ($^\circ$) \\
\hline

\multirow{2}{*}{Ferroelectric}
& $Cc$      & 6.020 & 6.020 & 12.937 & 94.466 & 94.466 & 119.977 \\
& $P31c$   & 6.016 & 6.016 & 12.696 & 90.000 & 90.000 & 120.000 \\
\hline

\multirow{2}{*}{Paraelectric}
& $C2/c$            & 5.977 & 5.977 & 12.758 & 94.436 & 94.436 & 119.989 \\
& $P\overline{3}1c$ & 5.981 & 5.981 & 12.444 & 90.000 & 90.000 & 120.000 \\
\hline

\multirow{2}{*}{Antiferroelectric}
& $P\overline{1}$   & 6.016 & 6.024 & 12.957 & 94.293 & 94.553 & 120.055 \\
& $P\overline{3}$   & 6.020 & 6.020 & 12.690 & 90.000 & 90.000 & 120.000 \\
\hline\hline
\end{tabular}
\end{table*}

\newpage

\section{Phonon Dispersions of All Phases}
\label{sec:phonons}

\begin{figure*}[htbp!]
\centering
\includegraphics[width=1\textwidth]{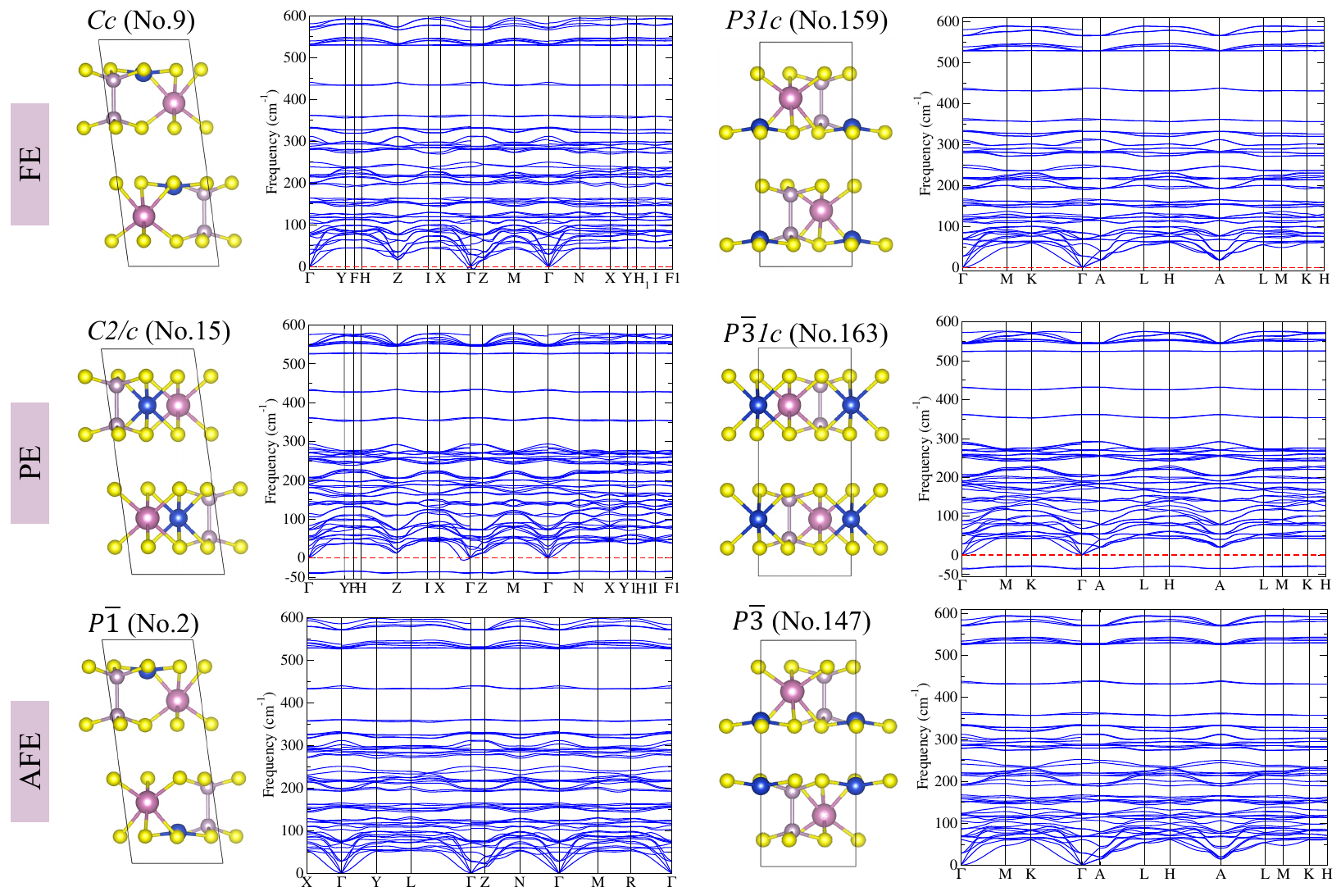}
\caption{\label{fig:phonon}
Crystal structures of all considered CuInP$_2$S$_6$ candidates in the ferroelectric (FE), paraelectric (PE), and antiferroelectric (AFE) phases, along with their corresponding calculated phonon dispersions including LO--TO splitting corrections for the polar phases at 0\,GPa.}
\end{figure*}

As discussed in the main text, the reference paraelectric $C2/c$ and $P\overline{3}1c$ phases exhibit two unstable phonon modes, which, upon following their instabilities, lead to the emergence of dynamically stable ferroelectric $Cc$ and $P31c$ phases, as well as antiferroelectric $P\overline{1}$ and $P\overline{3}$ phases, respectively. The phonon dispersions of all candidate structures are presented in this section, as shown in Fig.~\ref{fig:phonon}. Each crystal contains 20 atoms in the unit cell, corresponding to a total of 60 vibrational modes, among which three are acoustic modes and the remaining 57 are optical modes.

\subsection{Phonon Dispersion Calculations and Supercell Convergence}
\label{sec:phonon_convergence}

\begin{figure}[htbp]
\centering
\includegraphics[width=0.5\linewidth]{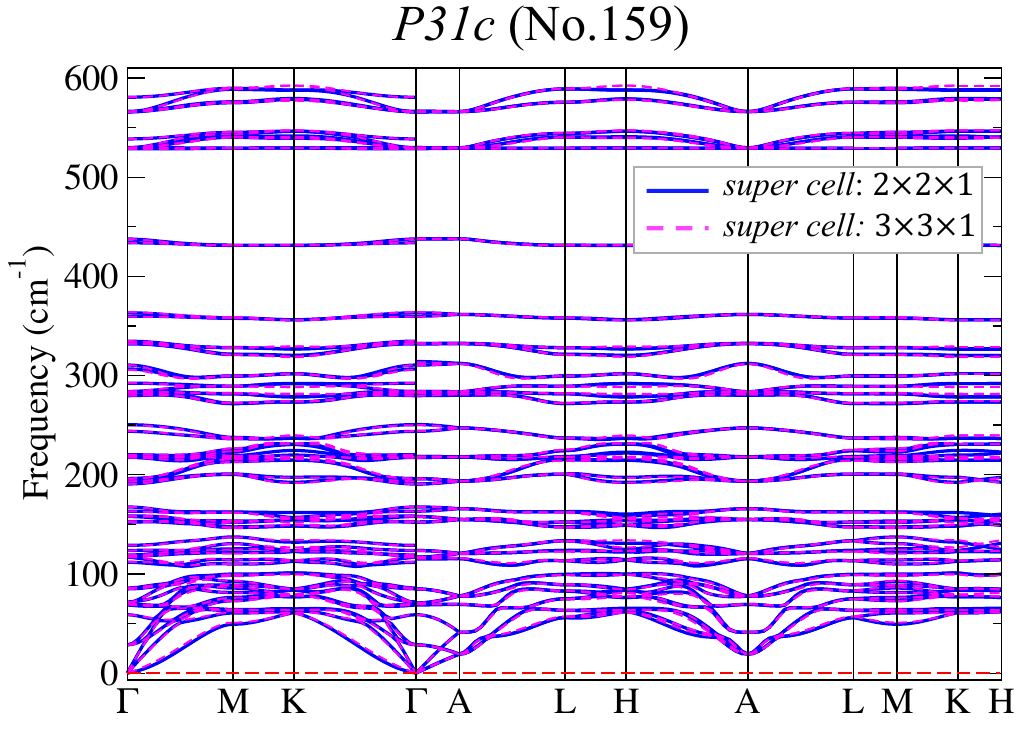}
\caption{\textcolor{black}{Comparison of the calculated phonon dispersions of the trigonal $P31c$ phase of CuInP$_2$S$_6$  using $2\times2\times1$ (solid lines) and $3\times3\times1$ (dashed lines) supercells within the finite-displacement method.}}
\label{fig:phonon_convergence}
\end{figure}

\textcolor{black}{Phonon dispersions were calculated using the finite-displacement method implemented in Phonopy. In this approach, the real-space harmonic interatomic force constants were obtained from DFT-calculated forces using $2\times2\times1$ supercells (VASP). The dynamical matrix at an arbitrary phonon wave vector $\mathbf{q}$ is constructed through the Fourier transformation of these force constants. 
To assess convergence with respect to supercell size, we additionally performed calculations using $3\times3\times1$ supercells for the trigonal $P31c$ phase. As shown in Fig.~\ref{fig:phonon_convergence}, the resulting phonon dispersions agree well with those obtained using $2\times2\times1$ supercells, indicating that the harmonic force constants are sufficiently converged with respect to the in-plane supercell dimensions. 
}

\section{Irreducible Representations of Paraelectric and Antiferroelectric Phases}
\label{sec:irreducible}

In this section, we present the allowed acoustic and optical vibrational modes at the Brillouin zone center for the paraelectric $C2/c$ and $P\overline{3}1c$ phases, as well as for the antiferroelectric $P\overline{1}$ and $P\overline{3}$ phases. The symmetry decomposition of these vibrational modes is obtained using group-theoretical analysis, and the corresponding irreducible representations are given below:

For the paraelectric $C2/c$ phase:
\begin{equation}
\begin{aligned}
\Gamma_{\text{acoustic}} &= A_u \oplus 2B_u, \\
\Gamma_{\text{optical}}  &= 14A_g \oplus 13A_u \oplus 16B_g \oplus 14B_u.
\end{aligned}
\label{eq:C2c_modes}
\end{equation}

For the paraelectric $P\overline{3}1c$ phase:
\begin{equation}
\begin{aligned}
\Gamma_{\text{acoustic}} &= A_{2u} \oplus E_u, \\
\Gamma_{\text{optical}}  &= 4A_{1g} \oplus 4A_{1u} \oplus 6A_{2g} \oplus 5A_{2u} \oplus 9E_u \oplus 10E_g.
\end{aligned}
\label{eq:P-31c_modes}
\end{equation}

For the antiferroelectric $P\overline{1}$ phase:
\begin{equation}
\begin{aligned}
\Gamma_{\text{acoustic}} &= 3A_u, \\
\Gamma_{\text{optical}}  &= 30A_g \oplus 27A_u.
\end{aligned}
\label{eq:P-1_modes}
\end{equation}

For the antiferroelectric $P\overline{3}$ phase:
\begin{equation}
\begin{aligned}
\Gamma_{\text{acoustic}} &= A_u \oplus {}^{1}E_u \oplus {}^{2}E_u, \\
\Gamma_{\text{optical}}  &= 10A_g \oplus 9A_u \oplus 10{}^{1}E_g \oplus 9{}^{1}E_u \oplus 10{}^{2}E_g \oplus 9{}^{2}E_u.
\end{aligned}
\label{eq:P-3_modes}
\end{equation}

\newpage

\section{Raman and IR Active Mode Frequencies for All CuInP$_2$S$_6$ Prototypes}
\label{sec:Raman_IR}

As discussed in the main text, the phonon dispersions with non-analytical term corrections (LO--TO splitting) have been evaluated for all considered CIPS prototypes, including the ferroelectric (FE), paraelectric (PE), and antiferroelectric (AFE) phases. In this section, we present the complete set of vibrational mode frequencies for each phase. Table\,\ref{tab:raman_modes} summarizes the Raman-active modes, while Table\,\ref{tab:IR_modes} lists the corresponding infrared-active mode frequencies.

\begin{table*}[htbp!]
\centering
\caption{Raman-active mode frequencies for different phases of CuInP$_2$S$_6$.}
\label{tab:raman_modes}
\renewcommand{\arraystretch}{1.35}
\setlength{\tabcolsep}{6pt}

\begin{tabular}{c c c p{0.72\textwidth}}
\hline\hline
\textbf{Type} & \textbf{Phase} & \textbf{Irrep} & \textbf{Frequencies (cm$^{-1}$)} \\
\hline

\multirow{6}{*}[-7.5em]{FE}
& \multirow{4}{*}{$Cc$}
& $A'$
& 29.3, 51.0, 74.0, 77.4, 104.2, 116.4, 119.1, 152.5, 154.2, 155.4, 163.7, 195.3, 201.2, 217.6, 249.3, 279.9, 282.3, 287.0, 300.6, 330.9, 331.6, 361.2, 433.5, 527.4, 530.9, 532.7, 565.5, 572.0 \\
\cline{3-4}
& & $A''$
& 28.5, 41.5, 73.0, 74.8, 81.4, 112.5, 117.5, 121.1, 153.2, 154.2, 164.3, 195.8, 200.5, 216.5, 219.4, 238.2, 279.5, 282.4, 287.6, 311.5, 331.0, 331.8, 361.3, 440.0, 529.4, 531.3, 531.8, 565.4, 572.3 \\
\cline{3-4}
& & $A'$(LO)
& 51.0, 104.2, 249.3, 279.9, 287.0, 300.6, 361.3, 433.5 \\
\cline{3-4}
& & $A'$(TO)
& 29.3, 74.0, 77.4, 116.4, 119.1, 152.5, 155.4, 163.8, 195.3, 201.2, 213.7, 217.6, 282.3, 330.9, 331.6, 527.4, 530.9, 532.7, 565.5, 572.0 \\

\cline{2-4}
& \multirow{2}{*}{$P31c$}
& $A_1$
& 69.6, 111.7, 167.5, 250.5, 280.0, 306.7, 359.7, 433.8, 528.6 \\
\cline{3-4}
& & $E$
& 28.1, 65.6, 83.2, 125.7, 129.3, 168.3, 175.4, 199.0, 208.2, 245.4, 252.6, 293.1, 299.1, 313.2, 315.6, 570.4, 572.4, 579.4, 581.1 \\
\hline

\multirow{4}{*}[-2.5em]{PE}
& \multirow{2}{*}{$C2/c$}
& $A_g$
& 22.2, 48.5, 73.3, 138.4, 181.8, 196.6, 208.0, 255.8, 261.8, 270.7, 361.2, 526.6, 546.1, 549.3 \\
\cline{3-4}
& & $B_g$
& 30.0, 56.7, 57.3, 75.0, 114.8, 139.7, 167.7, 188.3, 203.7, 254.5, 271.7, 293.5, 434.0, 547.3, 548.8 \\

\cline{2-4}
& \multirow{2}{*}{$P\overline{3}1c$}
& $A_{1g}$
& 190.8, 256.8, 361.3, 524.3 \\
\cline{3-4}
& & $E_g$
& 28.6, 56.3, 88.3, 141.7, 187.1, 206.6, 254.3, 268.9, 544.1, 546.9 \\
\hline

\multirow{3}{*}[-1.7em]{AFE}
& $P\overline{1}$
& $A_g$
& 27.3, 28.5, 49.9, 71.0, 74.4, 76.9, 112.1, 115.8, 122.7, 154.4, 157.9, 164.9, 192.8, 199.5, 216.4, 218.1, 251.1, 278.5, 284.6, 289.7, 311.8, 328.9, 330.4, 360.6, 440.0, 526.7, 529.5, 530.8, 571.9, 572.3 \\

\cline{2-4}
& \multirow{2}{*}{$P\overline{3}$}
& $A_g$
& 61.1, 68.3, 121.5, 167.5, 252.1, 278.9, 311.9, 359.4, 437.9, 527.7 \\
\cline{3-4}
& & $E_g$
& 29.3, 86.9, 123.7, 158.4, 193.0, 219.8, 285.5, 333.2, 526.7, 571.8 \\
\hline\hline
\end{tabular}
\end{table*}

\begin{table*}[htbp!]
\centering
\caption{IR-active mode frequencies for different phases of CuInP$_2$S$_6$.}
\label{tab:IR_modes}
\renewcommand{\arraystretch}{1.35}
\setlength{\tabcolsep}{6pt}

\begin{tabular}{c c c p{0.72\textwidth}}
\hline\hline
\textbf{Type} & \textbf{Phase} & \textbf{Irrep} & \textbf{Frequencies (cm$^{-1}$)} \\
\hline

\multirow{6}{*}[-7.5em]{FE}
& \multirow{4}{*}{$Cc$}
& $A'$
& 29.3, 51.0, 74.0, 77.4, 104.2, 116.4, 119.1, 152.5, 154.2, 155.4, 163.7, 195.3, 201.2, 217.6, 249.3, 279.9, 282.3, 287.0, 300.6, 330.9, 331.6, 361.2, 433.5, 527.4, 530.9, 532.7, 565.5, 572.0 \\
\cline{3-4}
& & $A''$
& 28.5, 41.5, 73.0, 74.8, 81.4, 112.5, 117.5, 121.1, 153.2, 154.2, 164.3, 195.8, 200.5, 216.5, 219.4, 238.2, 279.5, 282.4, 287.6, 311.5, 331.0, 331.8, 361.3, 440.0, 529.4, 531.3, 531.8, 565.4, 572.3 \\
\cline{3-4}
& & $A'$(LO)
& 51.0, 104.2, 249.3, 279.9, 287.0, 300.6, 361.3, 433.5 \\
\cline{3-4}
& & $A'$(TO)
& 29.3, 74.0, 77.4, 116.4, 119.1, 152.5, 155.4, 163.8, 195.3, 201.2, 213.7, 217.6, 282.3, 330.9, 331.6, 527.4, 530.9, 532.7, 565.5, 572.0 \\

\cline{2-4}
& \multirow{2}{*}{$P31c$}
& $A_1$
& 69.6, 111.7, 167.5, 250.5, 280.0, 306.7, 359.7, 433.8, 528.6 \\
\cline{3-4}
& & $E$
& 28.1, 65.6, 83.2, 125.7, 129.3, 168.3, 175.4, 199.0, 208.2, 245.4, 252.6, 293.1, 299.1, 313.2, 315.6, 570.4, 572.4, 579.4, 581.1 \\
\hline

\multirow{4}{*}[-2.5em]{PE}
& \multirow{2}{*}{$C2/c$}
& $A_u$
& 45.6, 71.7, 135.9, 181.4, 190.8, 202.9, 251.7, 255.4, 268.4, 360.2, 525.0, 545.2, 549.4  \\
\cline{3-4}
& & $B_u$
& 51.6, 72.9, 105.6, 136.2, 169.5, 185.8, 206.8, 257.2, 268.4, 283.1, 431.3, 545.8, 549.4  \\

\cline{2-4}
& \multirow{2}{*}{$P\overline{3}1c$}
& $A_{2u}$
& 107.3, 169.5, 287.2 \\
\cline{3-4}
& & $E_u$
& 53.1, 68.3, 134.7, 181.1, 201.1, 258.7, 271.1, 544.5, 545.6 \\
\hline

\multirow{3}{*}[-1.7em]{AFE}
& $P\overline{1}$
& $A_u$
& 60.4, 71.1, 73.2, 103.1, 115.1, 118.6, 153.7, 157.3, 160.2, 192.6, 195.8, 216.2, 217.4, 239.6, 277.2, 284.2, 289.4, 301.3, 327.7, 330.8, 360.5, 434.6, 526.7, 527.2, 532.0, 571.0, 572.3  \\

\cline{2-4}
& \multirow{2}{*}{$P\overline{3}$}
& $A_u$
& 69.5, 105.3, 162.8, 241.1, 278.4, 304.4, 363.3, 434.4, 527.1 \\
\cline{3-4}
& & $E_u$
& 68.4, 112.1, 151.0, 190.0, 214.1, 288.0, 332.9, 525.1, 569.4\\
\hline\hline
\end{tabular}
\end{table*}

\newpage

\section{\textcolor{black}{Raman and IR Intensities}}
\label{sec:Raman_IR_intensity}

\begin{figure}[htbp]
    \centering
    \includegraphics[width=0.98\linewidth]{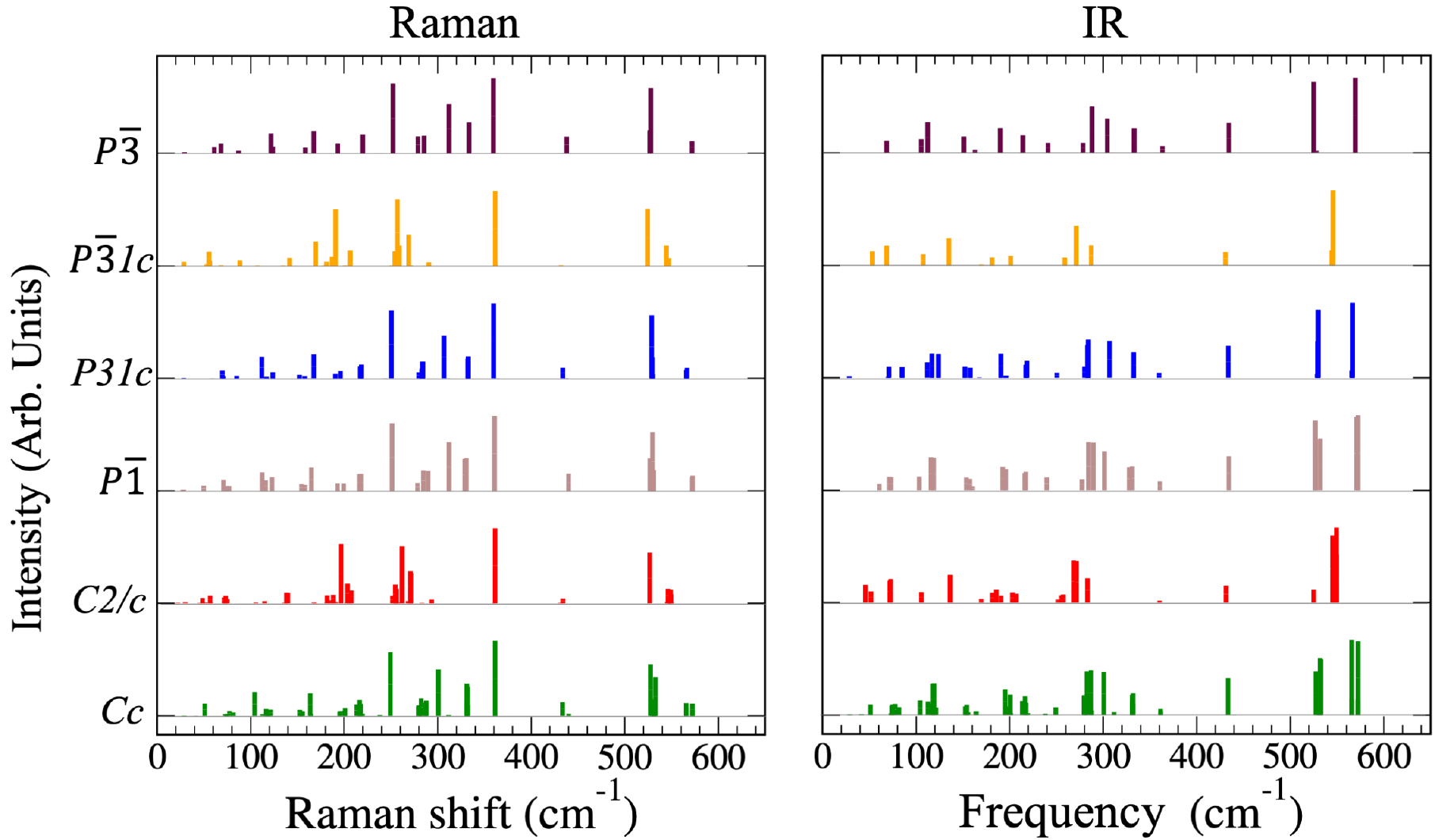}
\caption{\textcolor{black}{DFT-calculated (PBEsol+D3) Raman and infrared (IR) intensities for all the six studied phases of CuInP$_2$S$_6$.}}
    \label{fig:Raman_IR_intensity}
\end{figure}

\textcolor{black}{Figure~\ref{fig:Raman_IR_intensity} presents the calculated Raman and infrared (IR) intensities for all six candidate phases of CuInP$_2$S$_6$. Comparing the spectra provides additional insight into the relative visibility of the vibrational modes and complements their frequencies and symmetry assignments. In particular, the calculated intensities help distinguish strong and weak modes, providing references for experimental phase identification.}

\newpage

\textcolor{black}{\section{Characteristic Raman Mode Displacements}}
\label{sec:atomic_displacements}

\begin{figure}[htbp]
    \centering
    \includegraphics[width=0.98\linewidth]{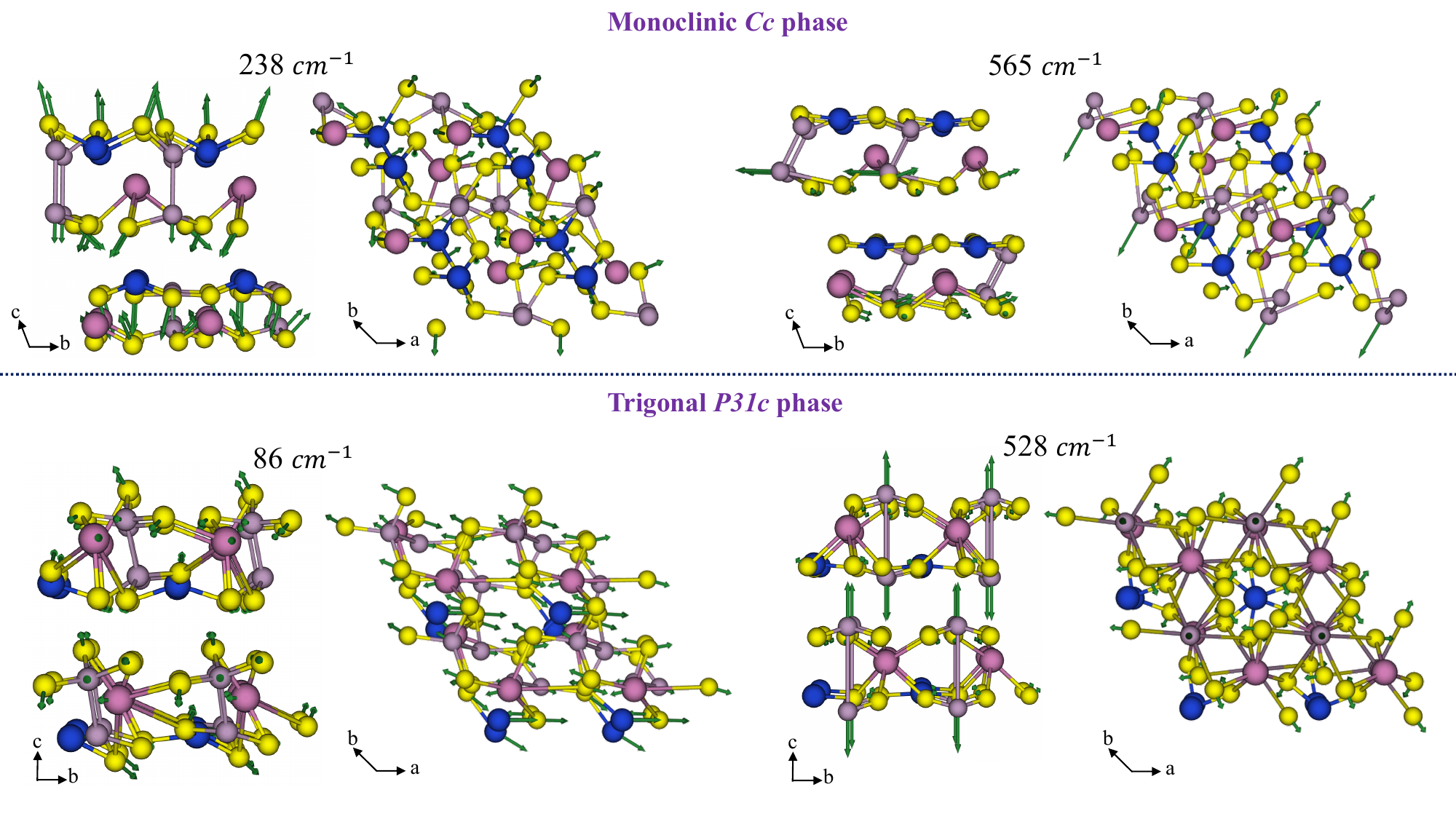}
    \caption{\textcolor{black}{Atomic displacement patterns of selected characteristic Raman-active phonon modes in CuInP$_2$S$_6$. Each mode is shown from two crystallographic viewing directions. Arrows indicate the relative atomic displacements associated with the phonon eigenvectors.}}
    \label{fig:signature_displacements}
\end{figure}

\textcolor{black}{The atomic displacement patterns of selected characteristic Raman modes for the monoclinic $Cc$ and trigonal $P31c$ phases are illustrated in Fig.~\ref{fig:signature_displacements}. For the monoclinic $Cc$ phase, the characteristic mode near 238~cm$^{-1}$ is dominated by out-of-plane S vibrations, whereas the high-frequency mode near 565~cm$^{-1}$ is primarily associated with in-plane P vibrations. For the trigonal $P31c$ phase, the low-frequency mode near 86~cm$^{-1}$ is primarily associated with in-plane vibrations of Cu and S atoms, while the strong mode near 528~cm$^{-1}$ involves out-of-plane P--P stretching coupled with in-plane S vibrations.  These displacement patterns provide insight into the atomic origins of the characteristic Raman fingerprints of the two polar phases.}

\newpage
\section{Polarization Selection Rules for the Studied Phases of CuInP$_2$S$_6$}
\label{sec:porto_notation}

As discussed in the main text, the polarization selection rules are determined by the Porto notation and the relative orientations of the incident and scattered light in the Raman experiment. Depending on the chosen scattering geometry, specific components of the Raman tensor are selected, which results in certain phonon modes being Raman active while others remain silent. 

In this section, we summarize all possible polarization configurations in backscattering geometry for the paraelectric, ferroelectric, and antiferroelectric phases of monoclinic and trigonal CuInP$_2$S$_6$ structures. The corresponding selection rules for each phase are presented in Table~\ref{tab:porto_supp_filled}.

\begin{table*}[htbp!]
\centering
\caption{Polarization selection rules for the studied phases of CuInP$_2$S$_6$ under additional Porto configurations.}
\label{tab:porto_supp_filled}

\renewcommand{\arraystretch}{1.25}
\setlength{\tabcolsep}{7pt}

\begin{tabular}{c c c c c c c}
\hline\hline

\diagbox[width=9em]{Geometry}{Phase}
& \textbf{Cc}
& \textbf{C2/c}
& $\boldsymbol{\mathrm{P\bar{1}}}$
& \textbf{P31c}
& $\boldsymbol{\mathrm{P\bar{3}1c}}$
& $\boldsymbol{\mathrm{P\bar{3}}}$ \\

\hline

$X(YY)\overline{X}$ 
& $A'(\mathrm{LO+TO})$
& $A_g$
& $A_g$
& $A_1(\mathrm{TO}),\,E(\mathrm{LO})$
& $A_{1g},\,E_g$
& $^{1}E_g,\,^{2}E_g,\,A_g$ \\

$X(YZ)\overline{X}$
& $A''(\mathrm{TO})$
& $B_g$
& $A_g$
& $E(\mathrm{TO})$
& $E_g$
& $^{1}E_g,\,^{2}E_g$ \\

$X(ZZ)\overline{X}$ 
& $A'(\mathrm{LO+TO})$
& $A_g$
& $A_g$
& $A_1(\mathrm{TO})$
& $A_{1g}$
& $A_g$ \\

$Y(XX)\overline{Y}$ 
& $A'(\mathrm{TO})$
& $A_g$
& $A_g$
& $A_1(\mathrm{TO}),\,E(\mathrm{TO})$
& $A_{1g},\,E_g$
& $^{1}E_g,\,^{2}E_g,\,A_g$ \\

$Y(XZ)\overline{Y}$
& \textcolor{black}{$A'(\mathrm{TO})$}
& \textcolor{black}{$A_g$}
& $A_g$
& $E(\mathrm{TO})$
& $E_g$
& $^{1}E_g,\,^{2}E_g$ \\

$Y(ZZ)\overline{Y}$ 
& $A'(\mathrm{TO})$
& $A_g$
& $A_g$
& $A_1(\mathrm{TO})$
& $A_{1g}$
& $A_g$ \\

\hline\hline
\end{tabular}
\end{table*}

% ===========================
% References
% ===========================
\vspace{6pt}

\vspace{6pt}